\documentclass[12pt,letterpaper]{article}
\usepackage{amsmath}
\usepackage{amsfonts}
\usepackage{amssymb}
\usepackage{rotating}
\usepackage{color}

\newtheorem{theo}{Theorem}

\newtheorem{defin}{Definition}
\newtheorem{prop}{Proposition}
\newtheorem{cor}{Corollary}
\newtheorem{rem}{Remark}

\newtheorem{ex}{Example}

\begin{document}

\title{When is tit-for-tat unbeatable?\thanks{
We are extremely thankful to two anonymous referees who provided very
valuable input. This paper subsumes the earlier working paper
\textquotedblleft Once beaten, never again: Imitation in two-player
potential games\textquotedblright . Some of the material was previously
circulated in a companion paper \textquotedblleft Unbeatable
Imitation\textquotedblright . We thank Carlos Al\'{o}s-Ferrer, Chen Bo, Drew
Fudenberg, Alexander Matros, Klaus Ritzberger, Karl Schlag, and John
Stachurski for interesting discussions. Seminar audiences at Australian
National University, Melbourne University, Monash University, UC Davis, UC
San Diego, the Universities of Heidelberg, Konstanz, Vienna, and Z\"{u}rich,
the University of Queensland, the University of Oregon, Calpoly, at the
International Conference on Game Theory in Stony Brook, 2009, the Midwestern
Economic Theory Conference in Evanston 2010, and at the Econometric Society
World Congress 2010 in Shanghai contributed helpful comments.}}
\author{Peter Duersch\thanks{%
Department of Economics, University of Heidelberg, Email:
peter.duersch@awi.uni-heidelberg.de} \and J\"{o}rg Oechssler \thanks{%
Department of Economics, University of Heidelberg, Email: oechssler@uni-hd.de%
} \and Burkhard C. Schipper\thanks{%
Department of Economics, University of California, Davis, Email:
bcschipper@ucdavis.edu} \and \medskip -- revised version --}
\maketitle

\begin{abstract}
\noindent We characterize the class of symmetric two-player games in which
tit-for-tat cannot be beaten even by very sophisticated opponents in a
repeated game. It turns out to be the class of exact potential games. More
generally, there is a class of simple imitation rules that includes
tit-for-tat but also imitate-the-best and imitate-if-better. Every decision
rule in this class is essentially unbeatable in exact potential games. Our
results apply to many interesting games including all symmetric 2x2 games,
and standard examples of Cournot duopoly, price competition, public goods
games, common pool resource games, and minimum effort coordination games.%
\newline


\noindent \textbf{JEL-Classifications: } C72, C73, D43.
\end{abstract}

\thispagestyle{empty}

\newpage \setcounter{page}{1}\setlength{\baselineskip}{1.5em}

\section{Introduction}

In a repeated two-player game, tit-for-tat refers to the strategy in which a
player always chooses the opponent's action from the previous round
(Rapoport and Chammah, 1965). Axelrod (1980a, 1980b) observed in his famous
tournaments that tit-for-tat when started with \textquotedblleft
cooperation\textquotedblright\ is surprisingly successful in sustaining
cooperation in the prisoners' dilemma. Part of the success is due to the
fact that it can hardly be exploited by an opponent who may follow various
more complex decision rules, a fact already noted in Axelrod (1980a).
Axelrod and Hamilton (1981) suggest that tit-for-tat is successful from an
evolutionary point of view (for a critique, see Selten and Hammerstein,
1984, or Nowak and Sigmund, 1993).

In this paper we extend the analysis to larger strategy sets and ask what is
the class of symmetric two-player games in which tit-for-tat cannot be
beaten by \emph{any} other decision rule? To be precise, suppose you play a
repeated symmetric game against one opponent. Suppose further that you know
that this opponent uses tit-for-tat. Since the rule is deterministic, you
therefore know exactly what your opponent will do in all future periods and
how he will react to your actions. The question we pose in this paper is
whether you can use this knowledge to exploit the tit-for-tat player in the
sense that you achieve a higher payoff than the tit-for-tat player. We call
tit-for-tat \emph{essentially unbeatable} if in the repeated game there
exists no strategy of the tit-for-tat player's opponent with which the
opponent can obtain, in total, over a possibly infinite number of periods, a
payoff difference that is more than the maximal payoff difference between
outcomes in the one--period game. It turns out that tit-for-tat is
essentially unbeatable if and only if the game is an exact potential game.
Exact potential games have been introduced by Monderer and Shapley (1996) to
show existence of and convergence to pure Nash equilibria. They are studied
widely in the literature on learning in games (see Sandholm, 2010). The
class of symmetric two-player exact potential games includes many meaningful
games such as all symmetric 2x2 games, and standard examples of Cournot
duopoly, price competition, public goods games, common pool resource games,
and minimum effort coordination games.

More generally, we show that there is a class of decision rules such that
any rule in this class is essentially unbeatable if the game is a two-player
symmetric exact potential game. This class of decision rules features
variants of imitation. It includes tit-for-tat but also imitate-the-best
(Vega-Redondo, 1997) and imitate-if-better (Duersch, Oechssler, and
Schipper, 2012a), where the latter prescribes to mimic the opponent's action
from the previous period if and only if the opponent received a higher
payoff in the previous period.

To gain some intuition for our results, consider the game of
\textquotedblleft chicken\textquotedblright\ presented in the following
payoff matrix.
\begin{equation*}
\begin{array}{cc}
&
\begin{array}{cc}
\text{swerve} & \text{straight}%
\end{array}
\\
\begin{array}{c}
\text{swerve} \\
\text{straight}%
\end{array}
& \left(
\begin{array}{cc}
3,3 & 1,4 \\
4,1 & 0,0%
\end{array}%
\right)%
\end{array}%
\end{equation*}%
What should a forward looking opponent do if she knows that she is facing a
tit-for-tat player?\footnote{%
In this paper we shall call the tit-for-tat player, or more generally, the
imitator \textquotedblleft he\textquotedblright\ and all possible opponents
\textquotedblleft she\textquotedblright .} To beat the tit-for-tat player
she can play \textquotedblleft straight\textquotedblright\ against
\textquotedblleft swerve\textquotedblright\ once. But if she wants to play
this again, she also has to play once \textquotedblleft
swerve\textquotedblright\ against \textquotedblleft
straight\textquotedblright , which equalizes the score. Thus, the maximum
payoff difference the opponent can obtain against a tit-for-tat player is 3,
the maximal one-period payoff differential.

Suppose now that the imitator uses the rule \textquotedblleft imitate if and
only if the other player obtained a higher payoff in the previous
round\textquotedblright\ and starts out with playing \textquotedblleft
swerve\textquotedblright . \ If the opponent decides to play
\textquotedblleft straight\textquotedblright , she will earn more than the
imitator today but will be copied by the imitator tomorrow. From then on,
the imitator will stay with \textquotedblleft straight\textquotedblright\
forever. If she decides to play \textquotedblleft swerve\textquotedblright\
today, then she will earn the same as the imitator and the imitator will
stay with \textquotedblleft swerve\textquotedblright\ as long as the
opponent stays with \textquotedblleft swerve\textquotedblright . Suppose the
opponent is a dynamic \emph{relative} payoff maximizer. In that case, the
dynamic relative payoff maximizer can beat the imitator at most by the
maximal one-period payoff differential of 3. Now suppose the opponent
maximizes the sum of her \emph{absolute} payoffs. The best an absolute
payoff maximizer can do is to play swerve forever. In this case the imitator
cannot be beaten at all as he receives the same payoff as his opponent. In
either case, imitation comes very close to the top--performing heuristics
and there is no evolutionary pressure against such a heuristic.

These results extend our recent paper Duersch, Oechssler, and Schipper
(2012a), in which we show that imitation is subject to a money pump (i.e.,
can be exploited without bounds) if and only if the relative payoff game is
of the rock-paper-scissors variety. The current results are stronger because
they show that imitation can only be exploited with a bound that is equal to
the payoff difference in the one--period game. Furthermore, the class of
imitation rules considered here is significantly broader. In particular, it
includes unconditional imitation rules like tit-for-tat. However, this comes
at the cost of having to restrict ourselves to the class of symmetric
two-player exact potential games. But as mentioned above, many economically
relevant games satisfy this property.

The behavior of learning heuristics such as imitation has previously been
studied mostly for the case when all players use the same heuristic. For the
case of imitate-the-best, Vega-Redondo (1997) showed that in a symmetric
Cournot oligopoly with imitators, the long run outcome converges to the
competitive output if small mistakes are allowed. This result has been
generalized to aggregative quasisubmodular games by Schipper (2003) and Al%
\'{o}s-Ferrer and Ania (2005). Huck, Normann, and Oechssler (1999),
Offerman, Potters, and Sonnemans (2002), and Apesteguia et al. (2007, 2010)
provide some experimental evidence in favor of imitative behavior. In
contrast to the above cited literature, the current paper deals with the
interaction of an imitator and a possibly forward looking, very rational and
patient player. Apart from experimental evidence in Duersch, Kolb,
Oechssler, and Schipper (2010) and our own paper Duersch, Oechssler, and
Schipper (2012a) we are not aware of any work that deals with this issue.
For a Cournot oligopoly with imitators and myopic best reply players,
Schipper (2009) showed that the imitators' long run average payoffs are
strictly higher than the best reply players' average payoffs.

The article is organized as follows. In the next section, we present the
model and provide a formal definition for being unbeatable. Section 3
introduces exact potential games. Necessary and sufficient conditions for
tit-for-tat to be essentially unbeatable are given in Section~\ref%
{essentially} followed by a number of potential applications. We finish with
Section~\ref{discussion}, where we summarize and discuss the results.

\section{Model}

We consider a symmetric two--player game $(X,\pi )$, in which both players
are endowed with the same compact set of pure actions $X$. For each player,
the continuous payoff function is denoted by $\pi :X\times X\longrightarrow
\mathbb{R}$, where $\pi(x, y)$ denotes the payoff to the player choosing the
first argument when his opponent chooses the second argument. We will
frequently make use of the following definition.

\begin{defin}[Relative payoff game]
\label{zerorel} Given a symmetric two-player game $(X, \pi)$, the relative
payoff game is $(X, \Delta)$, where the relative payoff function $\Delta: X
\times X \longrightarrow \mathbb{R}$ is defined by
\begin{equation*}
\Delta (x, y) = \pi(x, y) - \pi(y, x).
\end{equation*}
\end{defin}

Note that, by construction, every relative payoff game is a symmetric
zero-sum game since $\Delta (x,y)=-\Delta (y,x)$.

Next, we specify a number of different imitation rules.

\begin{defin}[Tit-for-tat]
A player plays strategy tit-for-tat if he plays in each period $t\geq 1$
whatever his opponent did in the preceding period $t-1$, i.e., $%
y_{t}=x_{t-1} $. Moreover, we allow $y_{0}$ to be arbitrary.\footnote{%
In Axelrod's tournament on the prisoners' dilemma, tit-for-tat submitted by
Anatol Rapoport also prescribed \textquotedblleft
cooperate\textquotedblright\ as initial action (see Axelrod, 1980a, 1980b).
While the prisoners' dilemma has a well-defined cooperative action, not all
games possess such an action. We therefore consider a definition of
tit-for-tat without restrictions on the initial action.}
\end{defin}

While tit-for-tat is an \emph{unconditional }imitation rule, in economics
imitation is often thought to be payoff dependent. The following decision
rule was studied by Vega-Redondo (1997). A player follows \emph{%
imitate-the-best} if for $t\geq 1$ his action is given by
\begin{equation*}
y_{t}\in \left\{
\begin{array}{cl}
\{x_{t-1}\} & \mbox{if }\Delta (x_{t-1},y_{t-1})>0 \\
\{x_{t-1},y_{t-1}\} & \mbox{if }\Delta (x_{t-1},y_{t-1})=0 \\
\{y_{t-1}\} & \mbox{if }\Delta (x_{t-1},y_{t-1})<0%
\end{array}%
\right.
\end{equation*}%
and arbitrary $y_{0}\in X$. Duersch, Oechssler, and Schipper (2012a) use a
version of it. A player follows \emph{imitate-if-better} if for $t\geq 1$
his action is given by
\begin{equation*}
y_{t}=\left\{
\begin{array}{cl}
x_{t-1} & \mbox{if }\Delta (x_{t-1},y_{t-1})>0 \\
y_{t-1} & \mbox{otherwise}%
\end{array}%
\right.
\end{equation*}%
and arbitrary $y_{0}\in X$.

The following class of imitation rules includes all three of the above
rules, tit-for-tat, imitate-the-best, and imitate-if-better.\footnote{%
In the context of the prisoner's dilemma, the class includes strategies like
two-tits-for-tat, \textquotedblleft always D\textquotedblright , and grimm
trigger but not tit-for-two-tats or Pavlov (Nowak and Sigmund, 1993).}

\begin{defin}[Imitation]
\label{imitation} We call a player an imitator if for $t\geq 1$, his action
is given by
\begin{equation*}
y_{t}\in \left\{
\begin{array}{cl}
\{x_{t-1}\} & \mbox{if }\Delta (x_{t-1},y_{t-1})>0 \\
\{x_{t-1},y_{t-1}\} & \mbox{otherwise}%
\end{array}%
\right.
\end{equation*}%
and arbitrary $y_{0}\in X$.
\end{defin}

That is, the imitator always adopts the opponent's action if in the previous
round the opponent's payoff was strictly higher than that of the imitator.
In other words, if the imitator decides to stick to his action, the other
player must have had a weakly lower payoff.

Our aim is to determine whether there exists a strategy with which the
imitator's opponent can obtain substantially higher payoffs than the
imitator. We allow for any strategy of the opponent, including very
sophisticated ones. In particular, the opponent may be infinitely patient
and forward looking, and may never make mistakes. More importantly, she may
know exactly what her opponent, the imitator, will do at all times,
including the imitator's starting value. She may also commit to any closed
loop strategy.

Consider now a situation in which the imitator starts out with a very
unfavorable initial action. A clever opponent who knows this initial action
can take advantage of it. Suppose that from then on the opponent has no
strategy that makes her better off than the imitator. Arguably, the
disadvantage in the initial period should not play a role in the long run.
This motivates the following definition.

\begin{defin}[Essentially unbeatable]
We say that imitation \emph{is essentially unbeatable} if for any strategy
of the opponent, the imitator can be beaten in total by at most the maximal
one-period payoff differential, i.e., if for any sequence of actions by the
opponent $(x_{0},x_{1},...)$ and any initial action $y_{0}$,
\begin{equation}
\sum_{t=0}^{T}\Delta (x_{t},y_{t})\leq \max_{x,y}\Delta (x,y),%
\mbox{ for
all }T\geq 0
\end{equation}%
where $y_{t}$ is defined by a specific imitation rule.
\end{defin}

\section{Exact potential games}

The question of unbeatability of imitation is closed linked to the class of
exact potential games (Monderer and Shapley, 1996).

\begin{defin}[Exact potential games]
\label{exact_pot} The symmetric game $(X,\pi )$ is an exact potential game
if there exists an exact potential function $P:X\times X\longrightarrow
\mathbb{R}$ such that for all $y\in X$ and all $x,x^{\prime }\in X$,%
\footnote{%
Given the symmetry of $(X,\pi )$, the second equation plays the role usually
played by the quantifier \textquotedblleft for all players\textquotedblleft\
in the definition of potential games.}
\begin{eqnarray*}
\pi (x,y)-\pi (x^{\prime },y) &=&P(x,y)-P(x^{\prime },y), \\
\pi (x,y)-\pi (x^{\prime },y) &=&P(y,x)-P(y,x^{\prime }).
\end{eqnarray*}
\end{defin}

Duersch, Oechssler, and Schipper (2012b) show that a symmetric two-player
game is an exact potential game if and only if its relative payoff game is
also an exact potential game. Furthermore, in symmetric two-player games,
exact potential games have a relative payoff function that is additively
separable, and has increasing and decreasing differences. All these
definitions may appear to be restrictive. However, we will show below that
there is a fairly large number of important examples that fall into this
class.

\begin{defin}[Additively separable]
A relative payoff function $\Delta $ is additively separable if $\Delta(x,y)
= f(x) + g(y)$ for some functions $f,g:X\longrightarrow \mathbb{R}$.
\end{defin}

Additive separable games have been studied in Balder (1997) and Peleg (1998).

\begin{defin}[Increasing/decreasing differences]
A (relative) payoff function $\Delta $ has decreasing (resp. increasing)
differences on $X\times X$ if there exists a total order $>$ on $X$ such
that for all $x^{\prime \prime },x^{\prime },y^{\prime \prime },y^{\prime
}\in X$ with $x^{\prime \prime }>x^{\prime }$ and $y^{\prime \prime
}>y^{\prime }$,
\begin{equation}
\Delta (x^{\prime \prime },y^{\prime \prime })-\Delta (x^{\prime },y^{\prime
\prime })\leq (\geq )\Delta (x^{\prime \prime },y^{\prime })-\Delta
(x^{\prime },y^{\prime }).  \label{decreasing diff}
\end{equation}
$\Delta $ is a valuation if it has both decreasing and increasing
differences.
\end{defin}

Games with increasing differences have been introduced by Topkis (1998).

It is interesting that in our context all of the above properties define the
same class of games.

\begin{rem}
\label{equivalence} For symmetric two-player games the following conditions
are equivalent: (i) $(X, \pi)$ is an exact potential game. (ii) $(X, \Delta
) $ is an exact potential game. (iii) $\Delta $ has increasing differences.
(iv) $\Delta$ has decreasing differences. (v) $\Delta $ is additively
separable.
\end{rem}

Duersch, Oechssler, and Schipper (2012b, Theorem 20) show that (i) and (ii)
are equivalent. There, we also show that (iii) and (iv) are equivalent for
all symmetric two-player zero-sum games (Proposition 13). Hence, (iii) or
(iv) imply that $\Delta $ is a valuation. Br\^{a}nzei, Mallozzi, and Tijs
(2003, Theorem 1) show that (ii) is equivalent to $\Delta $ being a
valuation for zero-sum games. Finally, Topkis (1998, Theorem 2.6.4.) shows
equivalence of (v) and $\Delta $ being a valuation for zero-sum games.

\section{Results\label{essentially}}

We are ready to state our main result.

\begin{theo}
\label{main} Let $(X, \pi)$ be a symmetric two-player game, where $X$ is
compact and $\pi$ is continuous. Tit-for-tat is essentially unbeatable if
and only if $(X, \pi)$ is an exact potential game.
\end{theo}

\noindent \textsc{Proof. } We prove here only ``$\Rightarrow$''. The
converse follows directly from Proposition~\ref{exactpot} below.

For tit-for-tat to be essentially unbeatable, for any $T > 0$ there must not
be a limit cycle $(x_{t}, y_{t})_{t = 0}^{T}$, with $\sum_{t =
0}^{T}\Delta(x_{t},y_{t}) > 0$ in which the tit-for-tat-player plays $y_{t}
= x_{t-1}$ for $t \geq 1$ and $y_0 = x_T$.

This implies in particular that for every 3-cycle $%
(x_{0},x_{2}),(x_{1},x_{0}),(x_{2},x_{1}),(x_{0},x_{2})...$ it must hold
that
\begin{equation}
\Delta (x_{0},x_{2})+\Delta (x_{1},x_{0})+\Delta (x_{2},x_{1})\leq 0.
\label{delta1}
\end{equation}%
Since this must hold for every 3-cycle, it must also hold for the reverse
cycle $(x_{2},x_{0}),(x_{1},x_{2}),\newline
(x_{0},x_{1}),(x_{2},x_{0}),...$, yielding
\begin{equation}
\Delta (x_{2},x_{0})+\Delta (x_{1},x_{2})+\Delta (x_{0},x_{1})\leq 0.
\label{delta2}
\end{equation}%
Since $(X,\Delta )$ is a symmetric zero-sum game, inequalities~(\ref{delta1}%
) and~(\ref{delta2}) imply
\begin{equation*}
\Delta (x_{0},x_{2})+\Delta (x_{1},x_{0})+\Delta (x_{2},x_{1})=0.
\end{equation*}

Hence,
\begin{equation*}
\Delta (x_{0},x_{2})+\Delta (x_{2},x_{1})=-\Delta (x_{1},x_{0})=\Delta
(x_{0},x_{1})
\end{equation*}%
and thus (since $\Delta (x,x)=0$ for any $x\in X$),
\begin{equation*}
\Delta (x_{0},x_{2})-\Delta (x_{2},x_{2})=\Delta (x_{0},x_{1})-\Delta
(x_{2},x_{1}).
\end{equation*}%
Since this equation holds for any 3-cycle, we have that $\Delta $ is a
valuation. I.e., for all $x^{\prime \prime },x^{\prime },x\in X,$
\begin{equation*}
\Delta (x^{\prime \prime },x)-\Delta (x^{\prime },x)=\Delta (x^{\prime
\prime },x^{\prime })-\Delta (x^{\prime },x^{\prime }).
\end{equation*}%
By Remark~\ref{equivalence} it implies that $(X,\pi )$ is an exact potential
game.\hfill $\Box $\newline

One direction of Theorem~\ref{main} actually holds for the more general
class of imitation rules given in Definition \ref{imitation}.

\begin{prop}
\label{exactpot} Let $(X,\pi )$ be a symmetric two-player game, where $X$ is
compact and $\pi $ is continuous. If $(X, \pi)$ is an exact potential game,
then any imitation rule in the class given in Definition~\ref{imitation} is
essentially unbeatable.
\end{prop}

\noindent \textsc{Proof. } By Remark~\ref{equivalence}, if $(X, \pi)$ is a
symmetric two-player exact potential game, then $\Delta$ is additively
separable. I.e., $\Delta(x,y) = f(x)+g(y)$ for some functions $f, g : X
\longrightarrow \mathbb{R}$. Since the relative payoff game is a symmetric
zero--sum game, we have that
\begin{equation*}
\Delta (x,x) = f(x) + g(x) = 0,
\end{equation*}
and hence
\begin{equation*}
\Delta (x,y)=f(x)-f(y).
\end{equation*}

Let $(x_{0},x_{1},...)$ be a sequence of actions generated by the opponent's
strategy, and let $\left\{\Delta (x_{t},y_{t})\right\}_{t = 0, 1, ...}$ be
her associated sequence of relative payoffs when the imitator follows an
imitation rule satisfying Definition~\ref{imitation}. We claim that
\begin{equation}
f(x_{t})-f(y_{t+1})\leq 0, \mbox{ for all } t \geq 0.  \label{f_diff}
\end{equation}
This follows because either the imitator imitates, i.e., $y_{t+1}=x_{t}$ or
he does not, $y_{t+1}=y_{t}$, in which case it must have been the case that $%
\Delta (x_{t},y_{t})=f(x_{t})-f(y_{t})\leq 0$.

Given (\ref{f_diff}), the sum of relative payoffs satisfies for any $T \geq
0 $,
\begin{equation*}
\sum_{t = 0}^{T} \Delta(x_{t},y_{t}) = \sum_{t=0}^{T} \left(f(x_{t}) -
f(y_{t})\right) \leq f(x_{T}) - f(y_{0}) = \Delta(x_{T},y_{0}) \leq
\max_{x,y}\Delta(x,y),
\end{equation*}
where $\max_{x,y} \Delta(x,y)$ exists because $\pi$ is continuous and $X$ is
compact.\hfill $\Box $\newline

In contrast to tit-for-tat, the existence of an exact potential function is
\emph{not} a necessary condition for being essentially unbeatable for
imitate-the-best and imitate-if-better. That is, going from tit-for-tat to
the more general class of imitation rules comes at the cost of losing the
converse of Theorem~\ref{main}. To see this, consider the following game,
which is not an exact potential game.
\begin{equation*}
\begin{array}{ccc}
\pi =%
\begin{array}{cc}
&
\begin{array}{ccccccc}
A &  &  & B &  &  & C%
\end{array}
\\
\begin{array}{c}
A \\
B \\
C%
\end{array}
& \left(
\begin{array}{ccc}
0,0 & 0,-1 & -1,0 \\
-1,0 & 0,0 & 0,10 \\
0,-1 & 10,0 & 0,0%
\end{array}%
\right)%
\end{array}
&  & \Delta =%
\begin{array}{cc}
&
\begin{array}{ccccc}
A &  & B &  & C%
\end{array}
\\
\begin{array}{c}
A \\
B \\
C%
\end{array}
& \left(
\begin{array}{ccc}
0 & 1 & -1 \\
-1 & 0 & -10 \\
1 & 10 & 0%
\end{array}%
\right)%
\end{array}%
\end{array}%
\end{equation*}%
It is easy to see that imitate-the-best and imitate-if-better are
essentially unbeatable for this game. However, tit--for--tat could be
exploited without any bound by following a cycle of actions $(A \rightarrow
B \rightarrow C \rightarrow A \dots )$. The reason for this difference is
that an imitate-the-best or imitate-if-better player would never leave
action $C$ whereas a tit--for--tat player can be induced to follow the
opponent from $C$ to $A$.

In the chicken game discussed in the Introduction, imitation was essentially
unbeatable since the maximal payoff difference was 3. Axelrod (1980a, b)
observed that tit-for-tat was unexploitable by other decision rules in the
prisoners' dilemma. More generally, since every symmetric $2\times 2$ is an
exact potential game, Proposition~\ref{exactpot} implies the following
corollary.

\begin{cor}
\label{2x2} In any symmetric 2x2 game, imitation is essentially unbeatable.
\end{cor}

It is also easy to see why the corollary is true without making use of the
notion of exact potential. Let $X=\{x,x^{\prime }\}$. Consider a period $t$
in which the opponent achieves a strictly positive relative payoff, $\Delta
(x,x^{\prime })>0$. (If no such period $t$ in which the opponent achieves a
strictly positive relative payoff exists, then trivially imitation is
essentially unbeatable.) The relative payoff game is symmetric zero-sum and
hence $\Delta (x^{\prime },x)=-\Delta (x,x^{\prime })$ and $\Delta
(x,x)=\Delta (x^{\prime },x^{\prime })=0$. Therefore, the action combination
$(x,x^{\prime })$ is the only one that makes imitation worse off. However,
in this case, imitation will immediately imitate and play $x$ in the next
round. The only way to move imitation back to play $x^{\prime }$ is for the
opponent to playing $x^{\prime }$ first, leading to the payoff $\Delta
(x^{\prime },x)=-\Delta (x,x^{\prime })<0$. Thus, every period with a
positive relative profit for the opponent must be preceded by a period with
a negative relative profit of the same absolute value. Depending on the
exact imitation rule, the imitator might not follow in the first period
where the opponent plays $x^{\prime }$, leading to even worse relative
profits, but it will always follow to $x$ right away.

Note that ``Matching pennies'' is not a counter-example since it is not
symmetric.

A sufficient condition for the additive separability of relative payoffs and
thus the existence of an exact potential for the symmetric two-player game $%
(X, \pi)$ is provided in the next result.

\begin{cor}
\label{separable2} Consider a game $(X, \pi)$ with a compact action set $X$
and a payoff function that can be written as $\pi(x,y)=f(x) + g(y) + a(x,y)$
for some continuous functions $f, g : X \longrightarrow \mathbb{R}$ and a
symmetric function $a: X \times X \longrightarrow \mathbb{R}$ (i.e., $%
a(x,y)=a(y,x)$ for all $x,y\in X$). Then imitation is essentially unbeatable.
\end{cor}

The following examples demonstrate that the assumption of additively
separable relative payoffs is not as restrictive as may be thought at first
glance.

\begin{ex}[Cournot Duopoly with Linear Demand]
\label{linear_Cournot} \emph{Consider a (quasi) Cournot duopoly given by the
symmetric payoff function }$\pi (x,y)=x(b-x-y)-c(x)$\emph{\ with }$b>0$.
\emph{Since }$\pi (x,y)$\emph{\ can be written as }$\pi
(x,y)=bx-x^{2}-c(x)-xy$\emph{, Corollary~\ref{separable2} applies, and
imitation is essentially unbeatable.}
\end{ex}

\begin{ex}[Bertrand Duopoly with Product Differentiation]
\label{Bertand} \emph{Consider a differentiated duopoly with constant
marginal costs, in which firms 1 and 2 set prices }$x$\emph{\ and }$y$\emph{%
, respectively. Firm 1's profit function is given by }$\pi (x,y)=(x-c)(a+by-%
\frac{1}{2}x)$\emph{, for }$a>0$\emph{, }$b\in \lbrack 0,1/2)$\emph{. Since }%
$\pi (x,y)$\emph{\ can be written as }$\pi (x,y)=ax-ac+\frac{1}{2}cx-\frac{1%
}{2}x^{2}-bcy+bxy$\emph{, Corollary~\ref{separable2} applies, and imitation
is essentially unbeatable. }
\end{ex}

\begin{ex}[Public Goods]
\emph{\label{public_goods} Consider the class of symmetric public good games
defined by }$\pi (x,y)=g(x,y)-c(x)$\emph{\ where }$g(x,y)$\emph{\ is some
symmetric monotone increasing benefit function and }$c(x)$\emph{\ is an
increasing cost function. Usually, it is assumed that }$g$\emph{\ is an
increasing function of the sum of provisions, }$x+y$\emph{. Various
assumptions on }$g$\emph{\ have been studied in the literature such as
increasing or decreasing returns. In any case, Corollary~\ref{separable2}
applies, and imitation is essentially unbeatable.}
\end{ex}

\begin{ex}[Common Pool Resources]
\emph{\label{CPR} Consider a common pool resource game with two
appropriators. Each appropriator has an endowment }$e>0$\emph{\ that can be
invested in an outside activity with marginal payoff }$c>0$\emph{\ or into
the common pool resource. Let }$x\in X\subseteq \lbrack 0,e]$\emph{\ denote
the opponent's investment into the common pool resource (likewise }$y$\emph{%
\ denotes the imitator's investment). The return from investment into the
common pool resource is }$\frac{x}{x+y}(a(x+y)-b(x+y)^{2})$\emph{, with }$%
a,b>0$\emph{. So the symmetric payoff function is given by }$\pi
(x,y)=c(e-x)+\frac{x}{x+y}(a(x+y)-b(x+y)^{2})$\emph{\ if }$x,y>0$\emph{\ and
}$ce$\emph{\ otherwise (see Walker, Gardner, and Ostrom, 1990). Since }$%
\Delta (x,y)=(c(e-x)+ax-bx^{2})-(c(e-y)+ay-by^{2})$\emph{, Proposition~\ref%
{exactpot} implies that imitation is essentially unbeatable.}
\end{ex}

\begin{ex}[Minimum Effort Coordination]
\emph{\label{minimum_effort} Consider the class of minimum effort games
given by the symmetric payoff function }$\pi (x,y)=\min \{x,y\}-c(x)$\emph{\
for some cost function }$c(\cdot )$ \emph{(see Bryant, 1983, and Van Huyck,
Battalio, and Beil, 1990). Corollary~\ref{separable2} implies that imitation
is essentially unbeatable.}
\end{ex}

\begin{ex}[Synergistic Relationship]
\emph{Consider a synergistic relationship among two individuals. If both
devote more effort to the relationship, then they are both better off, but
for any given effort of the opponent, the return of the player's effort
first increases and then decreases. The symmetric payoff function is given
by }$\pi (x,y)=x(c+y-x)$\emph{\ with }$c>0$\emph{\ and }$x,y\in X\subset
\mathbb{R}_{+}$\emph{\ with }$X$\emph{\ compact (see Osborne, 2004, p.39).
Corollary~\ref{separable2} implies that imitation is essentially unbeatable.}
\end{ex}

\begin{ex}[Diamond's Search]
\emph{\label{diamond_search} Consider two players who exert effort searching
for a trading partner. Any trader's probability of finding another
particular trader is proportional to his own effort and the effort by the
other. The payoff function is given by }$\pi (x,y)=\alpha xy-c(x)$ \emph{for}
$\alpha >0$ \emph{and} $c$ \emph{increasing (see Milgrom and Roberts, 1990,
p. 1270). The relative payoff game of this two-player game is additively
separable. By Proposition~\ref{exactpot} imitation is essentially unbeatable.%
}
\end{ex}

\section{Discussion\label{discussion}}

We have shown in this paper that there is a class of imitation rules that is
surprisingly robust to exploitation by any strategy in symmetric two-player
exact potential games. This includes strategies by truly sophisticated
opponents. The property that imitation is unbeatable in these games seems to
be unique among commonly used learning rules. We are not aware of other
rules outside our class of imitation rules share this property. For example,
many commonly used belief learning rules, such as best response learning or
fictitious play, can easily be exploited in all games in which a Stackelberg
leader achieves a higher payoff than the follower (as e.g. in Cournot
games). Against such rules, the opponent can simply stubbornly choose the
Stackelberg leader action knowing that the belief learning player will
eventually converge to the Stackelberg follower action. Thus, belief
learning rules can be beaten without bounds in such games. Yet, it remains
an open question for future research whether there are other behavioral
rules that perform equally well as imitation.

In Duersch, Oechssler, and Schipper (2012a) we show some extensions to more
general classes of two-player games including relative payoff games with a
generalized ordinal potential that come at cost of weakening the criterion
of essentially unbeatable and focusing on the rule ``imitate-if-better''.
Extensions to $n$-player games must be left for further research.

\end{document}